\documentclass[article,nojss]{jss}
\usepackage{orcidlink,thumbpdf,lmodern}
\graphicspath{{Figures/}}
\shortcites{imageio,numpy}

\author{
  Reto Stauffer~\orcidlink{0000-0002-3798-5507}\\Universit\"at Innsbruck \And
  Achim Zeileis~\orcidlink{0000-0003-0918-3766}\\Universit\"at Innsbruck
}
\Plainauthor{Reto Stauffer, Achim Zeileis}

\title{\pkg{colorspace}: A \proglang{Python} Toolbox for Manipulating and Assessing Colors and Palettes}
\Plaintitle{colorspace: A Python Toolbox for Manipulating and Assessing Colors and Palettes}
\Shorttitle{\pkg{colorspace}: Manipulating and Assessing Colors and Palettes in \proglang{Python}}

\Abstract{
  The \proglang{Python} \pkg{colorspace} package provides a toolbox for mapping
  between different color spaces which can then be used to generate a wide
  range of perceptually-based color palettes for qualitative or
  quantitative (sequential or diverging) information. These palettes (as
  well as any other sets of colors) can be visualized, assessed, and
  manipulated in various ways, e.g., by color swatches, emulating the
  effects of color vision deficiencies, or depicting the perceptual
  properties. Finally, the color palettes generated by the package can be
  easily integrated into standard visualization workflows in \proglang{Python}, e.g.,
  using \pkg{matplotlib}, \pkg{seaborn}, or \pkg{plotly}.
}
\Plainkeywords{Python, color palettes, color vision, visualization, assesment}

\Address{
  Reto Stauffer, Achim Zeileis\\
  Universit\"at Innsbruck\\
  Department of Statistics\\
  \emph{and}\\
  Digital Science Center\\
  6020 Innsbruck, Austria\\
  E-mail: \email{Reto.Stauffer@uibk.ac.at}, \email{Achim.Zeileis@R-project.org}\\
  URL: \url{https://retostauffer.org/}, \url{https://www.zeileis.org/}\\
}

\begin{document}

\vspace*{-0.5cm}

\section{Motivation}

Color is an integral element of visualizations and graphics and is
essential for communicating (scientific) information. However, colors
need to be chosen carefully so that they support the information
displayed for all viewers \citep[see e.g.,][]{Tufte:1990,Ware:2004,Wilke:2019}.
Therefore, suitable color palettes have been proposed in
the literature \citep[e.g.,][]{Brewer:1999,Ihaka:2003,Crameri:2020} and
many software packages transitioned to better color defaults over the
last decade. A prominent example from the \proglang{Python} community is
\pkg{matplotlib}~2.0 \citep{matplotlib20:colors} which replaced the
classic ``jet'' palette (a variation of the infamous ``rainbow'') by the
perceptually-based ``viridis'' palette. Hence a wide range of useful
palettes for different purposes is provided in a number of \proglang{Python}
packages today, including \pkg{cmcramery} \citep{cmcrameri},
\pkg{colormap} \citep{colormap}, \pkg{colormaps} \citep{colormaps},
\pkg{matplotlib} \citep{matplotlib}, \pkg{palettable}
\citep{palettable}, or \pkg{seaborn} \citep{seaborn}.

However, in most graphics packages colors are provided as a fixed set.
While this makes it easy to use them in
different applications, it is usually not easy to modify the perceptual
properties or to set up new palettes following the same principles. The
\pkg{colorspace} package addresses this by supporting color
descriptions using different color spaces (hence the package name),
including some that are based on human color perception. One notable
example is the Hue-Chroma-Luminance (HCL) model which represents colors
by coordinates on three perceptually-based axes: Hue (type of color),
chroma (colorfulness), and luminance (brightness). Selecting colors
along paths along these axes allows for intuitive construction of
palettes that closely match many of the palettes provided in the
packages listed above.

In addition to functions and interactive apps for HCL-based colors, the
\pkg{colorspace} package also offers functions and classes for
handling, transforming, and visualizing color palettes (from any
source). In particular, this includes the simulation of color vision
deficiencies \citep{Machado:2009} but also contrast ratios, desaturation,
lightening/darkening, etc.

The \pkg{colorspace} \proglang{Python} package was inspired by the eponymous R
package \citep{Zeileis:2020}. It comes with extensive documentation at
\url{https://retostauffer.github.io/python-colorspace/}, including many
practical examples. Selected highlights are presented in the following.

\section{Key functionality}

\subsection{HCL-based color palettes}

The key functions and classes for constructing color palettes using
hue-chroma-luminance paths (and then mapping these to hex codes) are:
\begin{itemize}
\item
  \code{qualitative_hcl}: For qualitative or unordered categorical
  information, where every color should receive a similar perceptual
  weight.
\item
  \code{sequential_hcl}: For ordered/numeric information from high to
  low (or vice versa).
\item
  \code{diverging_hcl}: For ordered/numeric information around a
  central neutral value, where colors diverge from neutral to two
  extremes.
\end{itemize}
These functions provide a range of named palettes inspired by
well-established packages but actually implemented using HCL paths.
Additionally, the HCL parameters can be modified or new palettes can be
created from scratch.

\begin{figure}[h!]
\centering
\includegraphics[width=0.7\textwidth]{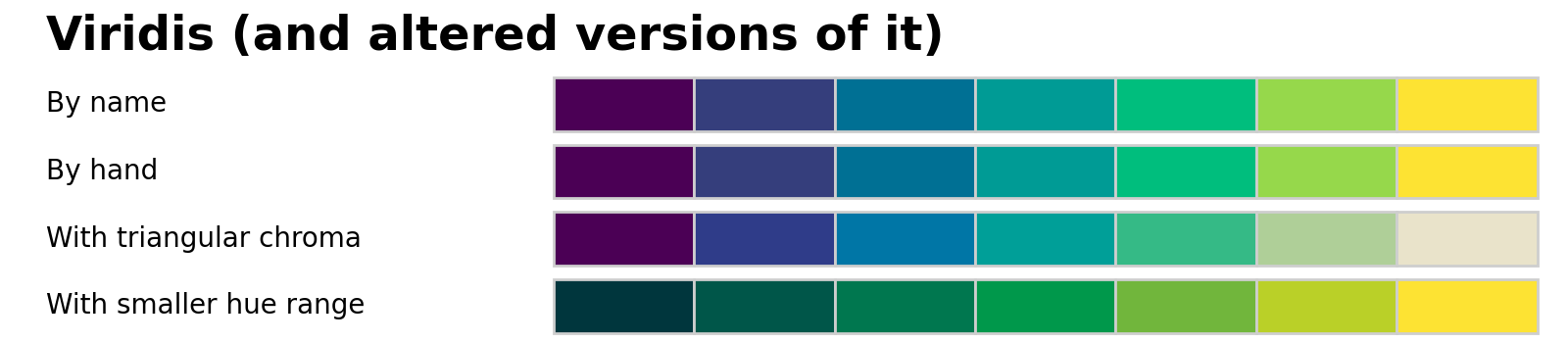}
\caption{Swatches of four HCL-based sequential palettes: \code{pal1}
is the predefined HCL-based viridis palette, \code{pal2} is identical
to \code{pal2} but created ``by hand'' and \code{pal3} and
\code{pal4} are modified versions with a triangular chroma paths and
reduced hue range, respectively.\label{fig:chosingpalettes}}
\end{figure}

As an example, Figure~\ref{fig:chosingpalettes} depicts color swatches for
four viridis variations. The first \code{pal1} sets up the palette
from its name. It is identical to the second \code{pal2} which
employes the HCL specification directly: The hue ranges from purple
(300) to yellow (75), colorfulness (chroma) increases from 40 to 95, and
luminance (brightness) from dark (15) to light (90). The \code{power}
parameter chooses a linear change in chroma and a slightly nonlinear
path for luminance.

In \code{pal3} and \code{pal4} the most HCL properties are kept the
same but some are modified: \code{pal3} uses a triangular chroma path
from 40 via 90 to 20, yielding muted colors at the end of the palette.
\code{pal4} just changes the starting hue for the palette to green
(200) instead of purple. All four palettes are visualized by the
\code{swatchplot} function from the package.

The objects returned by the palette functions provide a series of
methods, e.g., \code{pal1.settings} for displaying the HCL parameters,
\code{pal1(3)} for obtaining a number of hex colors, or
\code{pal1.cmap()} for setting up a \pkg{matplotlib} color map, among
others.

\begin{verbatim}
from colorspace import palette, sequential_hcl, swatchplot

pal1 = sequential_hcl(palette = "viridis")
pal2 = sequential_hcl(h = [300, 75], c = [40, 95], l = [15, 90],
                      power = [1., 1.1])
pal3 = sequential_hcl(palette = "viridis", cmax = 90,  c2 = 20)
pal4 = sequential_hcl(palette = "viridis", h1 = 200)

swatchplot({"Viridis (and altered versions of it)": [
               palette(pal1(7), "By name"), 
               palette(pal2(7), "By hand"),
               palette(pal3(7), "With triangular chroma"),
               palette(pal4(7), "With smaller hue range")
           ]}, figsize = (8, 1.75));
\end{verbatim}

\begin{figure}[h!]
\centering
\includegraphics[width=\textwidth]{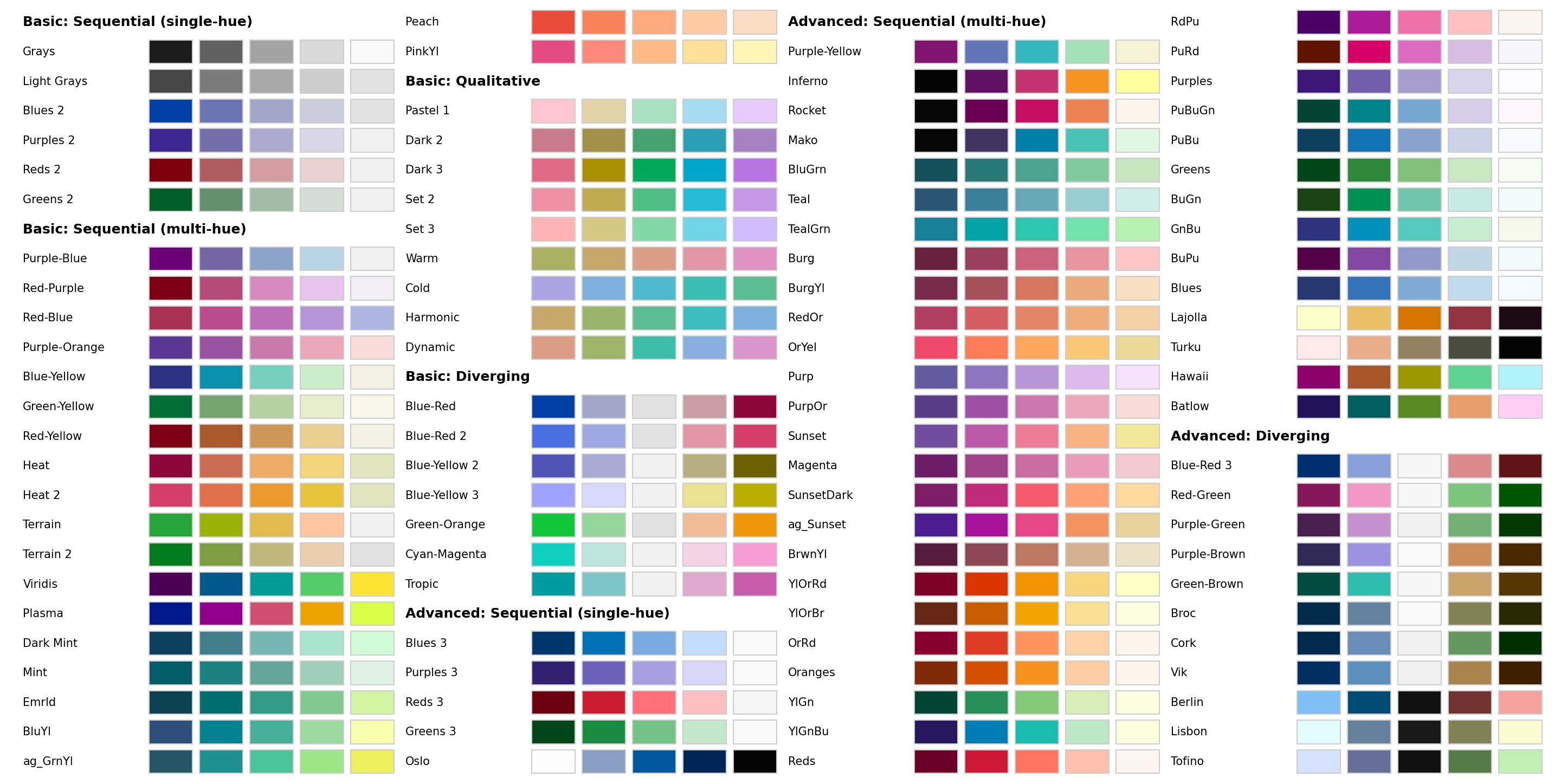}
\caption{Overview of the predefined (fully customizable) HCL color
palettes.\label{fig-hcl-palettes}}
\end{figure}

An overview of the named HCL-based palettes in \pkg{colorspace} is
depicted in Figure~\ref{fig-hcl-palettes}.

\begin{verbatim}
from colorspace import hcl_palettes
hcl_palettes(plot = True, figsize = (20, 15))
\end{verbatim}

\subsection{Palette visualization and assessment}

\begin{figure}[h!]
\centering
\includegraphics[width=\textwidth]{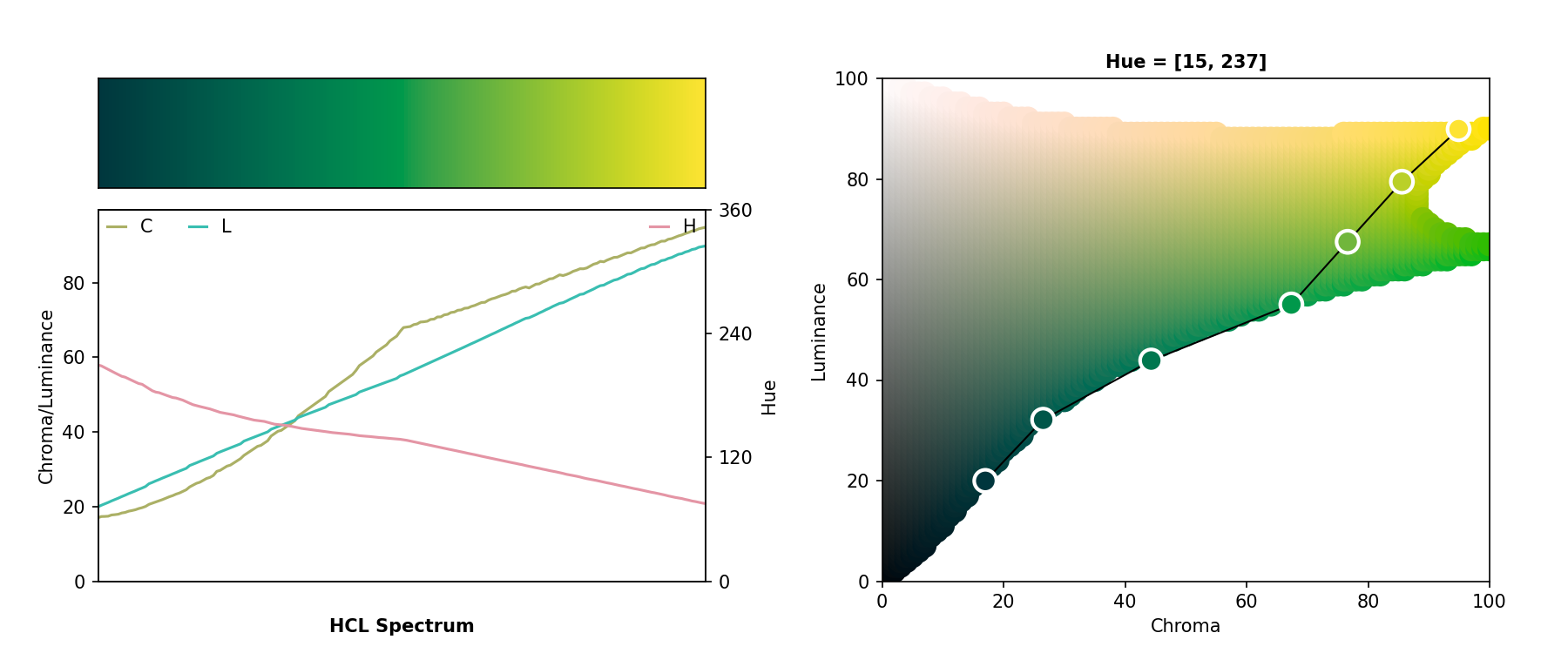}
\caption{Hue-chroma-luminance spectrum plot (left) and corresponding
path in the chroma-luminance coordinate system (where hue changes with
luminance) for the custom sequential palette
\code{pal4}.\label{fig:specplothclplot}}
\end{figure}

To better understand the properties of palette \code{pal4}, defined
above, Figure~\ref{fig:specplothclplot} shows its HCL spectrum (left) and
the corresponding path through the HCL space (right).

The spectrum in the first panel shows how the hue (right axis) changes
from about 200 (green) to 75 (yellow), while chroma and luminance (left
axis) increase from about 20 to 95. Note that the kink in the chroma
curve for the greenish colors occurs because such dark greens cannot
have higher chromas when represented through RGB-based hex codes. The
same is visible in the second panel where the path moves along the outer
edge of the HCL space.

\begin{verbatim}
pal4.specplot(figsize = (5, 5));
pal4.hclplot(n = 7, figsize = (5, 5));
\end{verbatim}

\subsection{Color vision deficiency}

\begin{figure}[h!]
\centering
\includegraphics[width=\textwidth]{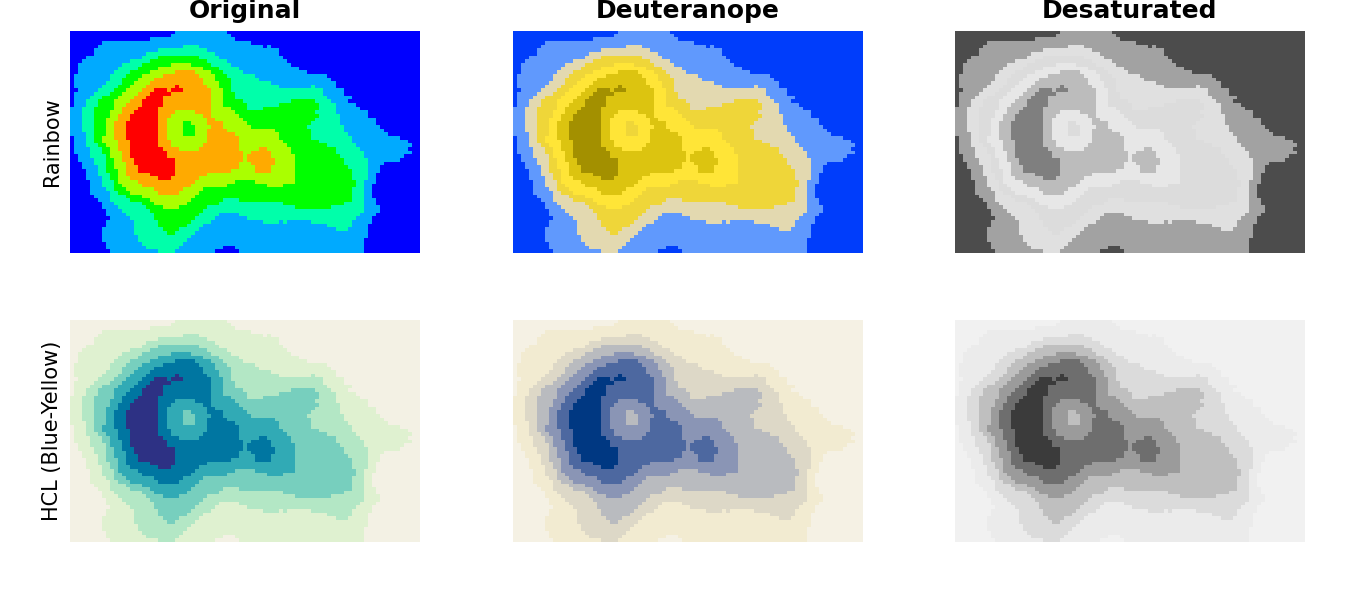}
\caption{Example of color vision deficiency emulation and color
manipulation using a heatmap. Top/bottom: RGB rainbow based palette and
HCL based sequential palette. Left to right: Original colors,
deuteranope color vision, and desaturated
representation.\label{fig-cvd}}
\end{figure}

Another important assessment of a color palette is how well it works for
viewers with color vision deficiencies. This is exemplified in
Figure~\ref{fig-cvd} depicting a demo plot (heatmap) under ``normal''
vision (left), deuteranomaly (colloquially known as ``red-green color
blindness'', center), and desaturated (gray scale, right). The palette
in the top row is the traditional fully-saturated RGB rainbow,
deliberately selected here as a palette with poor perceptual properties.
It is contrasted with a perceptually-based sequential blue-yellow HCL
palette in the bottom row.

The sequential HCL palette is monotonic in luminance so that it is easy
to distinguish high-density and low-density regions under deuteranomaly
and desaturation. However, the rainbow is non-monotonic in luminance and
parts of the red-green contrasts collapse under deuteranomaly, making it
much harder to interpret correctly.

\begin{verbatim}
from colorspace import rainbow, sequential_hcl
col1 = rainbow(end = 2/3, rev = True)(7)
col2 = sequential_hcl("Blue-Yellow", rev = True)(7)

from colorspace import demoplot, deutan, desaturate
import matplotlib.pyplot as plt

fig, ax = plt.subplots(2, 3, figsize = (9, 4))
demoplot(col1, "Heatmap", ax = ax[0,0], ylabel = "Rainbow", title = "Original")
demoplot(col2, "Heatmap", ax = ax[1,0], ylabel = "HCL (Blue-Yellow)")
demoplot(deutan(col1), "Heatmap", ax = ax[0,1], title = "Deuteranope")
demoplot(deutan(col2), "Heatmap", ax = ax[1,1])
demoplot(desaturate(col1), "Heatmap", ax = ax[0,2], title = "Desaturated")
demoplot(desaturate(col2), "Heatmap", ax = ax[1,2])
plt.show()
\end{verbatim}

\subsection[Integration with Python graphics packages]{Integration with \proglang{Python} graphics packages}

\begin{figure}[h!]
\centering
\includegraphics[width=\textwidth]{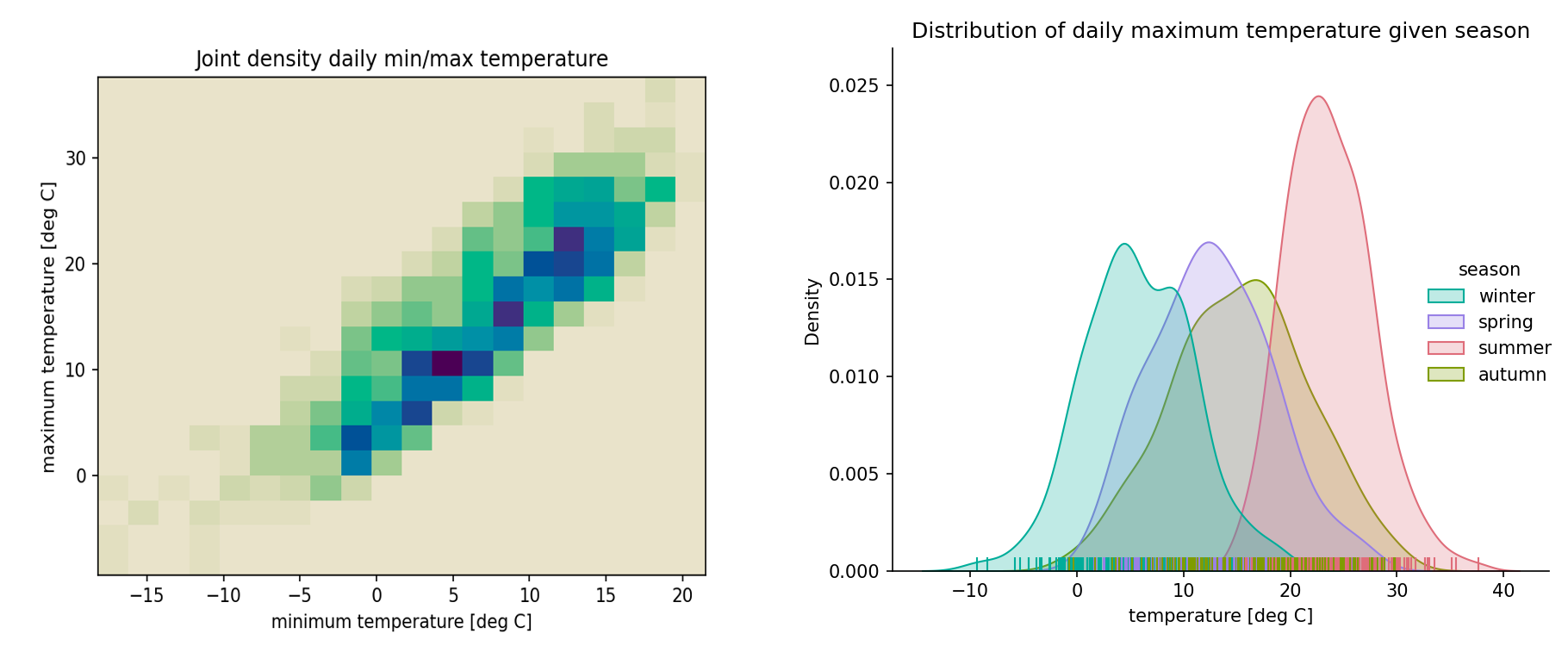}
\caption{Example of a \code{matplotlib} heatmap and a \code{seaborn}
density using custom HCL-based colors.\label{fig-plotting}}
\end{figure}

To illustrate that \pkg{colorspace} can be easily combined with
different graphics workflows in \proglang{Python}, Figure~\ref{fig-plotting} shows a
heatmap (two-dimensional histogram) from \pkg{matplotlib} and
multi-group density from \pkg{seaborn}. The code below employs an
example data set from the package (using \pkg{pandas}) with daily
maximum and minimum temperature. For \pkg{matplotlib} the colormap
(\code{.cmap()}; \code{LinearSegmentedColormap}) is extracted from
the adapted viridis palette \code{pal3} defined above. For
\pkg{seaborn} the hex codes from a custom qualitative palette are
extracted via \code{.colors(4)}.
\begin{verbatim}
from colorspace import dataset, qualitative_hcl
import matplotlib.pyplot as plt
import seaborn as sns

df = dataset("HarzTraffic")

fig = plt.hist2d(df.tempmin, df.tempmax, bins = 20,
                 cmap = pal3.cmap().reversed())
plt.title("Joint density daily min/max temperature")
plt.xlabel("minimum temperature [deg C]")
plt.ylabel("maximum temperature [deg C]")
plt.show()

pal = qualitative_hcl("Dark 3", h1 = -180, h2 = 100)
g = sns.displot(data = df, x = "tempmax", hue = "season", fill = "season",   
                kind = "kde", rug = True, height = 4, aspect = 1,
                palette = pal.colors(4))
g.set_axis_labels("temperature [deg C]")              
g.set(title = "Distribution of daily maximum temperature given season")
plt.show()
\end{verbatim}

\section{Dependencies and availability}

The \pkg{colorspace} is available from PyPI at
\url{https://pypi.org/project/colorspace}. It is designed to be
lightweight, requiring only \pkg{numpy} \citep{numpy} for the core
functionality. Only a few features rely on \pkg{matplotlib},
\pkg{imageio} \citep{imageio}, and \pkg{pandas} \citep{pandas}. More
information and an interactive interface can be found on
\url{https://hclwizard.org/}. Package development is hosted on GitHub at
\url{https://github.com/retostauffer/python-colorspace}. Bug reports,
code contributions, and feature requests are warmly welcome.

\bibliography{paper}

\end{document}